\documentclass[modern]{aastex61}

\def\msun{{\rm ~M}_{\odot}}

\def\mdot{\dot M}

\shorttitle{Model for Spectral States}

\shortauthors{Taam, et al.}

\begin{document}

\title{A Model for Spectral States and Their Transition in Cyg X-1}

\correspondingauthor{B.F Liu}

\email{bfliu@nao.cas.cn}

\author{Ronald E. Taam}
\affiliation{Institute of Astronomy and Astrophysics - TIARA, Academia Sinica, P.O. Box 23-141, Taipei, 10617, Taiwan}
\affiliation{Department of Physics and Astronomy, Northwestern University, 2145 Sheridan Road, Evanston, IL 60208, USA }

\author{Erlin Qiao}
\affiliation{Key Laboratory of Space Astronomy and Technology, National Astronomical Observatories, 
Chinese Academy of Sciences, Beijing 100012, China}
\affiliation{School of Astronomy and Space Science, University of Chinese Academy of Sciences, 19A 
Yuquan Road, Beijing 100049, China}

\author{B. F. Liu}
\affiliation{Key Laboratory of Space Astronomy and Technology, National Astronomical Observatories, 
Chinese Academy of Sciences, Beijing 100012, China}
\affiliation{School of Astronomy and Space Science, University of Chinese Academy of Sciences, 19A 
Yuquan Road, Beijing 100049, China}

\author{E. Meyer-Hofmeister}
\affiliation{Max-Planck-Institut f\"ur Astrophysik, Karl Schwarzschildstr. 1, D-85740, Garching, Germany}

\begin{abstract}

A new accretion picture based on a small disk surrounding a black hole is developed for the wind-fed source Cyg X-1.  The hard and soft spectral states of Cyg X-1 are interpreted in terms of co-spatial two component flows 
for the innermost region of an accretion disk. The state transitions result from the outward expansion and inward recession of this inner disk for the hard to soft and soft to hard transition respectively. The theoretical framework for state transitions in black hole X-ray binaries with high mass companions involving a change in the  inner disk size, thus, differs from systems with low mass companions involving the change in the outer disk size.  This fundamental difference stems from the fact that matter captured and supplied to the black hole  in wind-fed systems has low specific angular momentum and is hot essentially heated in the bow and spiral shocks, whereas it has high specific angular momentum and is cool in 
Roche lobe overflow systems.  The existence of a weak cool disk around  the ISCO region in the hard state  allows for the presence of a relativistically broadened Fe K line. The small disk fed by gas condensation 
forms without an extensive outer disk, precluding thermal instabilities and large outbursts, resulting in the lack of large amplitude outbursts and hysteresis effects in the light curve of high mass black hole X-ray binaries. Their  relatively persistent X-ray emission is attributed to their wind-fed nature.  
\end{abstract}
\keywords{accretion, accretion disks --- black hole physics ---
X-rays: stars --- X-rays: binaries --- stars:individual (Cyg X-1)}

\section{Introduction}

The nature of the X-ray source Cyg X-1 has attracted great interest since its discovery as a high mass X-ray binary system containing a black hole in a tight orbit.  The system is composed of a black hole of about $15 \msun$ and a massive companion of about $19 \msun$ orbiting each other with a period of 5.6 days. For a review of the 
parameters of the system, see the study by Orosz et al. (2011).  

The X-ray emission of Cyg X-1 is characterized by either a hard or soft spectral state and its temporal variations reveal transitions from a hard to a soft state and vice versa at accretion rates on the order of several hundredths of the Eddington value (e.g., Done et al. 2007 and references therein). In contrast to the case for black hole X-ray 
binaries with low mass companions, the luminosity variations do not exhibit large outbursts.  Specifically, the 
variations are limited to about a factor of 3 in Cyg X-1 over long time scales (Done \& Gierlinski 2003; Malyshev et al. 2013; Grinberg et al. 2014; Sugimoto et al. 2016; Zdziarski et al. 2017).  
In addition, there is no evidence of hysteresis in the light curve during the transitions. That is, the luminosity level at which Cyg X-1 exhibits a transition from a hard to a soft state is not significantly higher than the 
luminosity level corresponding to the transition from the soft to hard state.  

It has been suggested that the absence of such hysteresis effects in the light curve may be related to the high mass 
nature of the companion star (Maccarone \& Coppi 2003). Here, the black hole in Cyg X-1 is wind-fed rather than fed 
via Roche lobe overflow (RLOF). In such a case, the accretion disk is likely to be small in spatial extent (see Smith, 
Heindl, \& Swank 2002) and the mass flow variations limited to a rather narrow range.  Detailed multi-dimensional hydrodynamical calculations at high spatial resolution by Walder et al. (2014) reveal that the disk is, indeed, small due to the low specific angular momentum of the captured matter and due to the transfer of angular momentum by the action of a chain of shock structures as matter flows toward the black hole. As a further 
difference, the gas immediately captured by the black hole from the wind is not cool, having been heated by the shock waves in the system. 

Against this backdrop, there has been a long lasting debate on disk truncation in Cyg X-1 ever since the work of Esin et al. (1998), who modeled the  hard spectral state using an inner ADAF and a truncated outer disk rather than a disk extending inward to the innermost stable circular orbit (ISCO).  Subsequently, many investigations have sought 
observational evidence for  disk truncation (e.g., Parker et al. 2015; Basak et al. 2017), which result in different conclusions depending on the assumed continuum model. However, given the physical state of the gas in the Cyg X-1 system,  the applicability of a truncated cool outer disk and hot inner  model developed for systems in which the black hole is fed by RLOF may be questioned for a wind-fed system.  Alternatively, the accretion disk can be considered to be described by an inner cool disk with an outer hot accretion flow. This is similar to the inner disk structure envisioned for the intermediate hard state of the low mass black hole binary systems GX 339-4 and SWIFT J1753.5-0127 discussed by Taam et al. (2008) and also by Qiao \& Liu (2012) for GX 339-4 and Cyg X-1. In those systems where mass transfer is promoted via RLOF the region interior to a truncated outer cool disk consists of a hot accretion flow outside of a much smaller cool disk. Since the mass supply in Cyg X-1 is heated by shocks, the outer accretion flow is expected to remain hot throughout  approaching $10^8-10^9$ K in the innermost regions due to the gravitational energy release associated with accretion.  Such a description would be similar to an ADAF type structure for the mass accretion rates, characterizing the spectral states estimated in the system (see below). For the very innermost regions, the cool disk forms from the condensation of the hot coronal gas. Observational support for the inference of such an inner cool disk in Cyg X-1 is provided by the presence of broad Fe K lines (Duro et al. 2016) and a low temperature ($\sim 0.1 - 0.3$ keV) emission component in the hard state (Di Salvo et al. 2001; Duro et al. 2016). 

In this Letter, we apply the disk coronal condensation model (Liu et al. 2006, 2011) to Cyg X-1 to describe the hard spectral state as considered earlier (Qiao \& Liu 2012),  where an outer truncated disk was presumed. Here, we describe the hard state as a hot accretion flow supplied by a wind, with a weak inner disk fed by hot gas condensation.  As the wind supply rate increases, the cool inner disk becomes stronger as a consequence of efficient condensation, leading to a transition from a hard to soft state. In the proposed model, the inner disk expands during the transition from the hard to soft state and recedes during the transition from the 
soft to hard state. In the next section we briefly describe our model and present illustrative calculations of the disk spectrum in \S 3.  Finally, we summarize and conclude in the last section.  

\section{Model Description}

We consider the steady state structure of the innermost regions of a disk surrounding a non rotating black hole of 
$15 \msun$.  The theoretical framework of our two component accretion flow model is based on the interactions of a hot coronal gas with an underlying cool disk configuration,  which is similar in a general sense to a model developed by Chakrabarti \& Titarchuk (1995) in that it considers two gaseous components.  However, our model significantly differs in the detailed interaction processes and basic underlying assumptions (see Liu et al. 2017).  In particular, account is taken of the energy coupling through electron thermal conduction, inverse Compton scattering of disk photons, reprocessing and reflection of the irradiation, and the exchange of mass and enthalpy between the these two gaseous components. Such interactions can lead to either evaporation of gas from a cool disk into a corona or the condensation of gas from the corona to a cool disk.  An equilibrium is eventually established between the hot and cold accretion flow, which determines the description of the accretion flow in terms of a disk, an ADAF, or a corona lying above a disk.  The final steady geometry of the accretion flows depends not only on the mass supply rate, but also on the property of the gas as supplied from either RLOF or winds.

As the detailed interaction processes coupling the hot and cool disk material are similar for accretion from a wind and from RLOF, we adopt the model developed by Liu et al. (2006, 2007), Meyer et al. (2007), and Taam et al. (2008), which was applied to the so-called intermediate hard state of black hole X-ray binaries with low mass companions.  Of import was the realization that a hot region could spatially co-exist with an underlying cool disk in the innermost region, in contrast to an innermost region described by a pure ADAF. However, such an inner disk is weak and clearly distinguishable from an outer disk reaching inward to the ISCO (Cabanac et al. 2009).  We suggest that the condensation process for LMXBs  also occurs in Cyg X-1. 

 The difference in the description of the accretion process between Cyg X-1 and LMXBs reflects the nature and property of the mass supply. Specifically, in wind-fed black hole X-ray binaries the spatial extent and thermal state of the rotation supported accretion flow in the outer regions can significantly differ from the cool flow in the hard spectral state assumed in the truncated disk model (e.g., Done et al. 2007). As revealed by the three dimensional hydrodynamical simulations by Walder et al. (2014) the mass is captured from a stellar wind within a radius, $R_{\rm cap}$, as introduced in the Bondi-Hoyle description.  As such, gas is  heated 
at the bow shock to temperatures of $\sim 10^7$ K (Walder 2018), enveloping the black hole and spiral shocks in the wake region. Thus, it is expected that the gas 
captured and supplied to the disk surrounding a black hole is in a hot, rather than cool, physical state.  

 Such a hot gas supply results in similar spectral features as that in the hard/intermediate state of LMXBs. With an increase of the hot gas supply, the condensation can be very efficient, leading to a strong disk coexisting with the corona. In contrast to the soft state in LMXBs, the corona in Cyg X-1 is strong as hot gas is continuously supplied to the corona.  In addition, the absence of an outer cool disk excludes the existence of a  thermal instability arising from the ionization of hydrogen.  As a result, transient behavior associated with large outbursts is averted. Since the occasional  fluctuations in the wind supply cannot lead to a very low quiescent state, a weak inner disk is expected to survive in the hard state.  Such a picture provides support for the interpretation of an inner disk origin for the broad Fe lines and of little or no hysteresis in the light curve.

 The numerical results at high spatial resolution using up to 19 levels of refinement (Walder et al. 2014) also show that the specific angular momentum 
of the accreted matter is small and that additional angular momentum is removed from the gas by the action of spiral shocks. In particular, the shocks can extend close to the black hole ($\sim 214 R_S$, where $R_S$ is the Schwarzschild radius), where the flow can remain asymmetric and non Keplerian.  We note that the disk assumed in our model is axisymmetric and nearly Keplerian, 
and it is anticipated that the model disk is formed within this region. We adopt the condensation radius as the outer edge of the cool inner disk (see Liu et al. 2006, 2011; Taam et al. 2008). This radius has been illustrated as a function of the mass flow rate $\dot M$ and the $\alpha$ viscosity parameter in Liu et al. (2011) 
when the effects of Compton cooling and bremsstrahlung are taken into account. It increases with increasing accretion rate at fixed $\alpha$ and with decreasing values of $\alpha$ for fixed accretion rate (see Figs. 2 and 3 in Taam et al. 2008). 

The mass capture rate can be estimated in terms of the Bondi-Hoyle prescription. In particular, the capture radius is given by 
\begin{equation}
R_{\rm cap} = {2 G M \over v^2}
\end{equation}
where M is the mass of the black hole and $v$ is the velocity of the wind. Taking a black hole mass of $15 \msun$ and 
a velocity  in the range of 800 - 1000 km s$^{-1}$, $R_{\rm cap} = 3.9 - 6.2 \times 10^{11}$ cm.  The rate at which mass
is captured by the black hole, $\dot M_{\rm cap}$, can be estimated by the solid angle subtended by the captured region and is given in terms of the wind loss rate, $\dot M_{\rm wind}$, of the companion as 
\begin{equation}
\dot M_{\rm cap} = {\pi R_{\rm cap}^2 \over {4 \pi A^2}} \dot M_{\rm wind}
\end{equation}
where $A$ is the orbital separation of the system.  For $A=3\times 10^{12}$ cm, $\dot M_{\rm cap} = 0.004 - 0.01 \dot M_{\rm wind}$. If we adopt a wind loss rate of $2.5 \times 10^{-6} \msun$ yr$^{-1}$ (Hutchings 1976; Gies et al. 2003), the capture rate is $\sim 1 - 2.5 \times 10^{-8} \msun$ yr$^{-1}\sim$ 0.03 - 0.075 times the Eddington accretion rate  ($\dot M_{\rm Edd} \equiv {L_{\rm Edd}\over 0.1 c^2}= 1.40 \times 10^{18} {M\over \msun}{\rm \ g\ s^{-1}}$  by adopting radiation efficiency of 0.1 and  Eddington luminosity $L_{\rm Edd} ={4\pi GM cm_p\over \sigma_T}=1.26 \times 10^{38} {M\over \msun}{\rm \ erg\ s^{-1}}$ for pure hydrogen)
which is taken as a range of the mass supply rate to accretion.

\section{Spectral Calculations} 

The detailed disk and corona structure envisioned here is similar to that considered in the context of active 
galactic nuclei (AGN), having been calculated in Liu et al.  (2015) and Qiao \& Liu (2017).  As illustrated in 
Figs. 2 and 4 of Liu et al (2015), the mass flow rate in the corona decreases with decreasing distance from the 
black hole, while the mass flow rate in the cool disk increases with decreasing distance. The geometrically thin 
disk retreats toward the ISCO with decreasing gas supply rate, leading to a spectral transition from a soft to hard 
state at a few percent of the Eddington accretion rate.  

Analogous to the AGN case, the structure and spectrum have been calculated for a range of accretion rates for a 
black hole mass of $15 \msun$ corresponding to Cyg X-1. The spatial variation of the mass flow rates and  the spectral energy distributions are respectively illustrated in Fig. 1 and Fig. 2 for two representative cases. A mass supply rate of $\dot M = 0.05 \dot M_{\rm Edd}$  is chosen for the first case, taking into account a luminosity of $\sim 0.05$ times the Eddington value in its soft state for Cyg X-1.  The second case represents a hard state, in which the hot accretion flow dominates over the thin disk.  Since the  radiation efficiency is low in an ADAF, while the luminosity in the hard state is not significantly smaller than that in the soft state for Cyg X-1, a mass supply rate of $\dot M = 0.04 \dot M_{\rm Edd}$ is adopted.  For illustrative purposes, $\alpha$=0.3(0.5), and albedo $a=$ 0.15(0.5) are adopted for 
the first (second) case.  In this example, more coronal radiation is expected to be reflected in the hard state as a 
consequence of less absorption by the weak disk.

Fig. 1 illustrates that the disk recedes to $34R_s$ for a mass supply rate $\dot M = 0.04 \dot M_{\rm Edd}$, while it is much larger for $\dot M = 0.05 \dot M_{\rm Edd}$ supported by gas condensation.  
The mass flow rate in the cool disk is much less  by about a factor of 10 than that in the corona at the ISCO region for the second case. 
This compares to the first case where the mass flow rate in the innermost region is comparable in the corona 
and cool disk.  It is clearly evident from Fig. 2 that the case for $\dot M = 0.05 \dot M_{\rm Edd}$ is softer than for the case in which $\dot M = 0.04 \dot M_{\rm Edd}$.  We identify the latter case 
to the hard spectral state and the former to the soft spectral state of Cyg X-1.  We note that the hard spectrum is primarily produced by an ADAF, but with a disk-like component corresponding to a maximal effective temperature of 0.19 keV, which contributes $ 30\%$ of the total luminosity.  In contrast, the disk component contributes $82\%$ of the total luminosity with an effective temperature of 0.32 keV at the soft state.  The bolometric luminosities for these two cases are  $1.0 \times10^{38} {\rm erg \  s^{-1}}$ ($=0.053L_{\rm Edd}$) and $3.0 \times10^{37} {\rm erg \  s^{-1}}$ ($=0.016L_{\rm Edd}$)   respectively,   yielding a ratio of 3.3.

Here, the state transition from the hard state to the soft state is attributed to the increase in the spatial extent 
of the inner cool disk reflecting the change in condensation radius.  Similarly, the state transition from the 
soft state to the hard state occurs as a result of a decrease in the condensation radius. 

   Very recently, the geometry of the accretion flows in black hole X-ray binaries in the hard spectral state has been investigated based on the interaction of hot and cold accretion flows (Poutanen et al. 2017). An inner ADAF connected to a truncated disk is excluded due to the problem of seed-photon starving. A static corona above a full cool disk is also ruled out since the spectrum is found to be too soft caused by the reprocessing of coronal emission. Such a conclusion is also drawn by our proposed model.  To reduce the reprocessing component, Poutanen et al. (2017) propose a truncated disk coexisting with hot gas either overlapping the disk or containing some cold gas within it.  This is in contrast to a small disk embedded in the innermost region of the hot accretion flow as a consequence of weak condensation of wind-fed accretion flow in the hard state proposed here. These favorable geometries are obtained by different approaches. Specifically, we consider both the energy and mass coupling between the disk and corona, calculating the mass flow rates in the cold and hot accretion flows, and the resulting temperature, density and spectra. Poutanen et al (2017) also take into account the energy balance, but focus on fitting to observational photon index by assuming the proper scattering optical depth and covering factor. Such models provide alternatives for interpreting the spectra observed in black hole X-ray binaries.

\section{Summary and Conclusion}

A new picture for the accretion in Cyg X-1 is proposed based on the formation of a small disk in X-ray sources powered by the gravitational energy release associated with wind-fed accretion.  The hard and soft spectral states of Cyg X-1 can, in principle, be understood in terms of co-spatial two component flows in which the coronal region overlies an optically thick region in the innermost accretion flows. Such a model  can be preferable to a truncation disk model (optically thick outer disk and optically thin inner ADAF) for the hard spectral state since the physical state of the captured matter from a high mass companion has undergone heating in the bow and spiral shocks  in addition to that resulting from the standard accretion processes that release energy as gas flows toward the black hole. The state transition from a hard state to the soft state is interpreted in terms of the outward expansion of an inner optically thick disk instead of the inward movement of a truncated optically thick outer disk.  During the soft state both models rely on the existence of an optically thick inner disk.  In our model the transition from the soft state to the hard state corresponds to the recession of the inner optically thick disk to a smaller spatial  extent, as a consequence of a decrease in the gas condensation at low accretion rates. 

 With the above interpretion for the spectral states of Cyg X-1, the application of the cool inner disk model is expanded beyond the interpretation for the intermediate state of black hole low mass X-ray binaries.  The model cannot be applied to the very low hard state of these systems since a permanent inner cool disk is absent at the very low luminosities corresponding to their quiescent state.  The absence of such a quiescent state in Cyg X-1 is critical for the existence of the cool disk in its hard state.

As a result, the soft X-ray emission that has been attributed to the emission from the inner edge of a truncated outer disk (e.g., Done et al.  2007) originates in the weak cool disk extending to the ISCO. As such, the model permits the presence of a relativistically broadened Fe K line in both spectral states. 
 
The small radii of the disk in the proposed model, resulting from the low specific angular momentum of the captured
matter, naturally leads to the absence of an extensive outer disk, precluding 
the possibility of thermal instabilities in the hydrogen ionization zone in an outer disk and, therefore, large 
outbursts. Luminosity fluctuations would still be present in wind-fed sources, perhaps as a consequence of turbulence 
in the accretion wake region (Walder et al. 2014), but these fluctuations are likely smaller than those due to thermal 
instabilities in the disk in sources fed by the mass transfer promoted by RLOF. Thus, the wind-fed sources could be  described as persistent. 
 
Given this picture, the interpretative framework for state transitions between black hole X-ray binaries with high 
mass companions (wind-fed systems) significantly differs from those with low mass companions (fed by RLOF).
In particular, such an interpretation  could be the key difference for the absence of hysteresis effects in the light curve between transient sources and persistent sources  (e.g. Maccarone \& Coppi 2003; Zdziarski \& Gierlinski  2004).  As illustrated in our earlier work (Meyer-Hofmeister et al. 2005; Liu et al. 2005), the hysteresis between the transition luminosities of hard-to-soft and soft-to-hard is caused by irradiation 
(Compton heating or cooling) before the state transition, as transient sources were in a deep quiescent state 
before outbursts are triggered. This effect is insignificant in persistent sources because of a small difference 
in luminosities in the hard and soft states.  Therefore, the light curve during the transition between states 
is likely to be more symmetrical in persistent systems.

Although we have related the nature of the companion star to the black hole as important for the interpretation of the 
cause of the spectral state transitions in Cyg X-1, the interpretation of the timing properties within the framework of 
our model remains challenging.  Studies of the time variability as inferred from the power density spectra (see Pottschmidt 
et al. 2003) reveal that the frequencies of the broad noise components based on a multiple Lorentzian profiles, where it 
can fit, increase as the Cyg X-1 softens. This suggests that the spatial region involved in producing these noise components 
is more compact with increasing luminosity, which is speculated to be associated with the variation of the truncation 
radius of the thin disk with the accretion rate.  In addition, the observed hard time lags initially discovered by Priedhorsky 
et al. (1979) and Nolan et al. (1981) and later described in more detail by Nowak et al. (1999) provide a further observational 
constraint that must be considered. To address this timing constraint, Kotov, Churazov, \& Gilfanov (2001) and more recently 
Rapisarda, Ingram, \& van der Klis (2017) developed a model for the hard lags based on the propagation of perturbations in 
the accretion flow in the framework of the truncated disk geometry developed for RLOF systems. In this interpretation, accretion 
rate fluctuations near the truncation radius of the disk propagate toward the black hole, leading to a lag of hard X-rays 
provided that the spectrum emitted from smaller radii is harder.  Their results raise difficulties with our proposed 
model for the spectral variability as it fails to explain the hard lags in its current framework.  However, it may
be possible that the cool inner disk underlying the hot corona is inhomogeneous in structure.  In such a case, a model 
involving cool blobs where the soft photons can introduce variability in the hard photons within the ADAF itself, similar 
to that developed by Bottcher \& Liang (1999) for the hard time lags, may be applicable.  Alternatively, a magnetic flare 
model for the hard lags as proposed by Poutanen \& Fabian (1999) involving Comptonization occurring in the corona above 
the inner cool disk may have some bearing on the lag issue. 

In the future, it is desirable to study the simplified model further to confront not only the timing data, but
also to understand the narrow luminosity range over which the state transition takes place, given the significant differences 
in radiative efficiencies between ADAF and optically thick accretion disk models. In particular, studies for the apparent 
differences in the $\alpha$ viscosity parameter and albedo in the two different spectral states are required. Finally, an 
understanding of the variations of the mass supply to the inner accretion disk surrounding the black hole would be needed 
for detailed comparison of observed light curves with the model. In this case, coupling of the inner flow to the outer 
flow determined by the gas capture process would be necessary.  Such studies would provide the critical insight necesssary 
to allow one to interpret the light curves and, hence, to constrain the fundamental parameters (i.e., mass accretion rate 
and $\alpha$ parameter) underlying the theoretical model. 

\acknowledgments

We thank the anonymous referee for his/her assessments on the model based on the observational perspective. Financial support for this work is provided by the National Program on Key Research and Development Project (Grant No. 
2016YFA0400804) and  the National Natural Science Foundation of China (grants 11673026 and 11773037). In addition, 
R.E.T. acknowledges support from the Theoretical Institute for Advanced Research in Astrophysics in the Academia Sinica 
Institute of Astronomy \& Astrophysics.
  

\clearpage

\begin{figure}[h]
\plotone{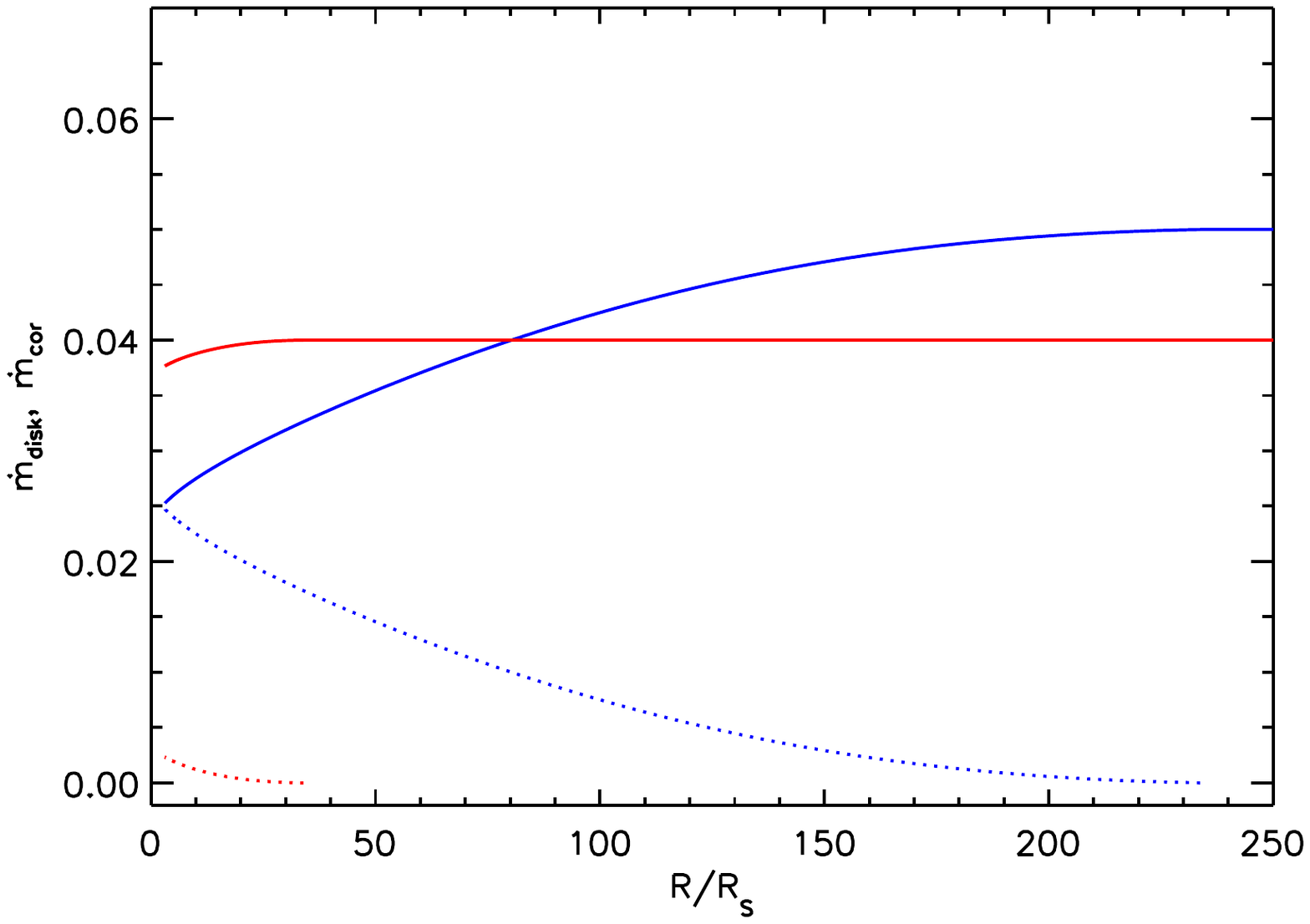}
\caption{The radial distribution of Eddington-scaled mass flow rate in the corona ($\dot m_{\rm cor}$, solid lines) and in the disk ($\dot m_{\rm disk}$, dotted lines)  for mass supply rates  of  $\dot M = 0.05 \dot M_{\rm Edd}$ (blue lines) and $\dot M = 0.04 \dot M_{\rm Edd}$ (red lines).  }
\end{figure} 

\clearpage

\begin{figure}[h]
\plotone{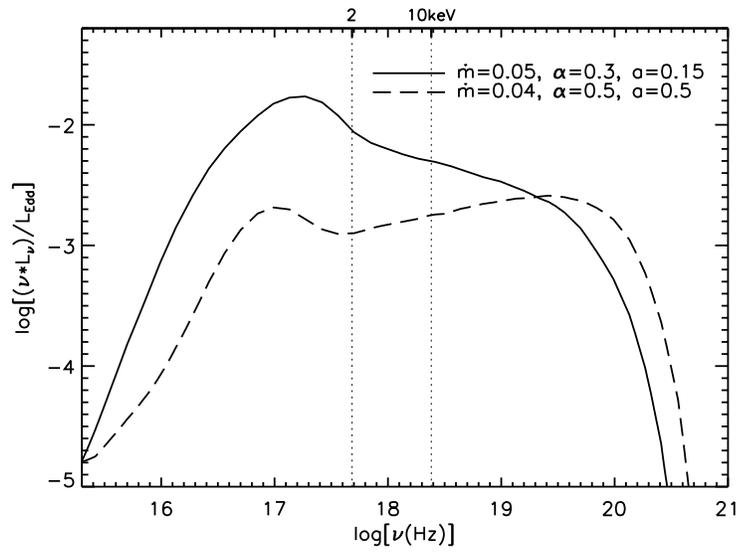}
\caption{The soft (solid lines) and hard ( dashed lines) spectra for Cyg X-1, where the accretion rates are $\dot m={\mdot  / \mdot_{\rm Edd}}=0.05$ and $0.04 $, respectively.}
\end{figure} 

\end{document}